\documentclass[journal]{IEEEtran}

\usepackage[pdftex]{graphicx}
\usepackage[cmex10]{amsmath}
\usepackage{bm}
\usepackage{cite}
\usepackage{textcomp}
\usepackage{SIunits}
\usepackage{todonotes}
\usepackage{multirow}
\usepackage[acronym]{glossaries} 
\usepackage{overpic}
\usepackage{hyperref}

\usepackage{xcolor}
\usepackage{color}
\definecolor{light}{rgb}{.1,.6,.1}

\makeglossaries 

\def\deg{{^\circ}}
\def\vec#1{\ensuremath{\bm{{#1}}}}

\DeclareMathOperator*{\argmax}{\arg\!\max}

\newglossaryentry{ibm}{name={IBM},description={ideal binary mask},
first={\glsentrydesc{ibm} (\glsentrytext{ibm})},
plural={IBMs},
descriptionplural={ideal binary masks},
firstplural={\glsentrydescplural{ibm} (\glsentryplural{ibm})}}
\newglossaryentry{itd}{name={ITD},description={interaural time difference},
first={\glsentrydesc{itd} (\glsentrytext{itd})},
plural={ITDs},
descriptionplural={interaural time differences},
firstplural={\glsentrydescplural{itd} (\glsentryplural{itd})}}
\newglossaryentry{ild}{name={ILD},description={interaural level difference},
first={\glsentrydesc{ild} (\glsentrytext{ild})},
plural={ILDs},
descriptionplural={interaural level differences},
firstplural={\glsentrydescplural{ild} (\glsentryplural{ild})}}
\newacronym{2d}{2D}{two-dimensional}
\newacronym{tf}{T-F}{time-frequency}
\newacronym{snr}{SNR}{signal-to-noise ratio}
\newacronym{ams}{AMS}{amplitude modulation spectrogram}
\newacronym{asa}{ASA}{auditory scene analysis}
\newacronym{erb}{ERB}{equivalent rectangular bandwidth}
\newacronym{ccf}{CCF}{cross-correlation function}
\newacronym{casa}{CASA}{computational auditory scene analysis}
\newglossaryentry{prior}{name=\emph{a priori},description={}}
\newglossaryentry{post}{name=\emph{a posteriori},description={}}
\newglossaryentry{hfa}{name=HIT\,-\,FA,description={}}
\newacronym{dnn}{DNN}{deep neural network}
\newacronym{rasta}{RASTA-PLP}{relative spectral transform and perceptual linear prediction}
\newacronym{mfcc}{MFCCs}{mel-frequency cepstral coefficients}
\newacronym{spp}{SPP}{speech presence probability}
\newacronym{drr}{DRR}{direct-to-reverberant ratio}
\newacronym{hats}{HATS}{head and torso simulator}
\newglossaryentry{brir}{name={BRIR},description={binaural room impulse response},
first={\glsentrydesc{brir} (\glsentrytext{brir})},
plural={BRIRs},
descriptionplural={binaural room impulse responses},
firstplural={\glsentrydescplural{brir} (\glsentryplural{brir})}}
\newglossaryentry{hrtf}{name={HRTF},description={head related transfer function},
first={\glsentrydesc{hrtf} (\glsentrytext{hrtf})},
plural={HRTFs},
descriptionplural={head related transfer functions},
firstplural={\glsentrydescplural{hrtf} (\glsentryplural{hrtf})}}
\newglossaryentry{hrir}{name={HRIR},description={head related impulse response},
first={\glsentrydesc{hrir} (\glsentrytext{hrir})},
plural={HRIRs},
descriptionplural={head related impulse responses},
firstplural={\glsentrydescplural{hrir} (\glsentryplural{hrir})}}
\newacronym{psd}{PSD}{power spectral density}
\newacronym{gmm}{GMM}{Gaussian mixture model}
\newacronym{kemar}{KEMAR}{Knowles Electronic Manikin for Acoustic Research}
\newacronym{lc}{LC}{local criterion}
\newacronym{rms}{RMS}{root mean square}
\newacronym{rmse}{RMSE}{root mean square error}
\newacronym{pdf}{PDF}{probability density function}
\newacronym{dft}{DFT}{discrete Fourier transform}
\newacronym{stft}{STFT}{short-time discrete Fourier transform}
\newacronym{em}{EM}{expectation-maximization}
\newacronym{mct}{MCT}{multi-conditional training}
\newacronym{ml}{ML}{most likely}
\newacronym{maa}{MAA}{minimum audible angle}

\begin{document}
%
\title{Exploiting Deep Neural Networks and Head Movements for Robust Binaural Localisation of Multiple Sources in Reverberant Environments}
%
%
%
\author{Ning Ma,
            Tobias May,
            and Guy J. Brown
\thanks{Ning Ma and Guy J. Brown are with the Department of Computer Science, University of Sheffield, Sheffield S1 4DP, UK (email: \{n.ma, g.j.brown\}@sheffield.ac.uk)}
\thanks{Tobias May is with Hearing Systems Group, Technical University of Denmark, DK - 2800 Kgs. Lyngby, Denmark (email: tobmay@elektro.dtu.dk)}
}

\markboth{IEEE/ACM Transactions on Audio, Speech and Language Processing}%
{Ma \MakeLowercase{\textit{et al.}}: Exploiting deep neural networks and head movements for robust binaural localisation of multiple sources in reverberant environments}




\maketitle



\begin{abstract}
This paper presents a novel machine-hearing system that exploits deep neural networks (DNNs) and head movements for robust binaural localisation of multiple sources in reverberant environments. DNNs are used to learn the relationship between the source azimuth and binaural cues, consisting of the complete cross-correlation function (CCF) and interaural level differences (ILDs). In contrast to many previous binaural hearing systems, the proposed approach is not restricted to localisation of sound sources in the frontal hemifield. Due to the similarity of binaural cues in the frontal and rear hemifields, front-back confusions often occur. To address this, a head movement strategy is incorporated in the localisation model to help reduce the front-back errors. The proposed DNN system is compared to a Gaussian mixture model (GMM) based system that employs interaural time differences (ITDs) and ILDs as localisation features. Our experiments show that the DNN is able to exploit information in the CCF that is not available in the ITD cue, which together with head movements substantially improves localisation accuracies under challenging acoustic scenarios in which multiple talkers and room reverberation are present.
\end{abstract}


\IEEEpeerreviewmaketitle

\glsresetall 

\vspace{.5cm}



\section{Introduction}
\label{s:intro}

%
%
%
%


\IEEEPARstart{T}{his} paper aims to reduce the gap in performance between human and machine sound localisation, in conditions where multiple sound sources and room reverberation are present. Human listeners have little difficulty in localising sounds under such conditions; they are able to decode the complex acoustic mixture that arrives at each ear with apparent ease~\cite{Blauert97}. In contrast, sound localisation by machine systems is usually unreliable in the presence of interfering sources and reverberation. This is the case even when an array of multiple microphones is employed~\cite{NadiriRafaely2014}, as opposed to the two (binaural) sensors available to human listeners.

The human auditory system determines the azimuth of sounds in the horizontal plane by using two principal cues: \glspl{itd} and \glspl{ild}. A number of authors have proposed binaural sound localisation systems that use the same approach, by extracting \glspl{itd} and \glspl{ild} from acoustic recordings made at each ear of an artificial head~\cite{WillertEtAl2006, MayvandeParKohlrausch11a, WoodruffWang12, MayMaBrown2015}. Typically, these systems first use a bank of cochlear filters to split the incoming sound into a number of frequency bands. The \gls{itd} and \gls{ild} are then estimated in each band, and statistical models such as \glspl{gmm} are used to determine the source azimuth from the corresponding binaural cues~\cite{MayMaBrown2015}. Furthermore, the robustness of this approach to varying acoustic conditions can be improved by using \gls{mct}. This introduces uncertainty into the statistical models of the binaural cues, enabling them to handle the effects of reverberation and interfering sound sources~\cite{MayvandeParKohlrausch11a,WoodruffWang12,MayMaBrown2015,MayvandeParKohlrausch13}.

In contrast to many previous machine systems, the approach proposed here is not restricted to sound localisation in the frontal hemifield; we consider source positions in the 360$\deg$ azimuth range around the head. In this unconstrained case, the location of a sound cannot be uniquely determined by \glspl{itd} and \glspl{ild}; due to the similarity of these cues in the frontal and rear hemifields, front-back confusions occur~\cite{WightmannKistler99}. Although machine listening studies have noted this as a problem \cite{MayMaBrown2015,MaEtAl2015}, listeners rarely make such confusions because head movements, as well as spectral cues due to the pinnae, play an important role in resolving front-back confusions~\cite{wallach1940,WightmannKistler99,McAnallyMartin14}.

Relatively few machine localisation systems have attempted to incorporate head movements. Braasch~\textit{et al.}~\cite{Braasch2013} averaged cross-correlation patterns across different head orientations in order to resolve front-back confusions in anechoic conditions. More recently, May~\textit{et al.}~\cite{MayMaBrown2015} combined head movements and \gls{mct} in a system that achieved robust sound localisation performance in reverberant conditions. In their approach, the localisation system included a hypothesis-driven feedback stage which triggered a head movement when the azimuth could not be unambiguously estimated. Subsequently, Ma \textit{et al.}~\cite{MaEtAl2015} evaluated the effectiveness of different head movement strategies, using a complex acoustic environment that included multiple sources and room reverberation. In agreement with studies on human sound localisation~\cite{Per1997}, they found that localisation errors were minimised by a strategy that rotated the head towards the target sound source.

This paper describes a novel machine-hearing system that robustly localises multiple talkers in reverberant environments, by combining \gls{dnn} classifiers and head movements. Recently, \glspl{dnn} have been shown to give state-of-the-art performance in a variety of speech recognition and acoustic signal processing tasks \cite{BengioYoshua2009}. In this study, we use \glspl{dnn} to map binaural features, obtained from an auditory model, to the corresponding source azimuth. Within each frequency band, a \glspl{dnn} takes as input features the \gls{ccf} (as opposed to a single estimate of \gls{itd}) and the \gls{ild}. Using the whole cross-correlation function provides the classifier with rich information for classifying the azimuth of the sound source~\cite{MaEtAl2015dnn}. A similar approach was used by \cite{JiangEtAl2014} and \cite{Yu2016} in binaural speech segregation systems. However, neither study specifically addressed source localisation because it was assumed that the target source was fixed at zero degrees azimuth.


The proposed binaural sound localisation system is described in detail in Section~\ref{s:system}. Section~\ref{s:eval} describes the evaluation framework and presents a number of source localisation experiments, in which head movements are simulated by using \glspl{brir} to generate direction-dependent binaural sound mixtures. Localisation results are presented in Section~\ref{s:expr}, which compares our DNN-based approach to a baseline method that uses \glspl{gmm}, and assesses the contribution that various components make to performance. The paper concludes with Section~\ref{s:conc}, which proposes some avenues for future research.


\section{System}
\label{s:system}

\begin{figure}
\centerline{\includegraphics[width=0.9\columnwidth]{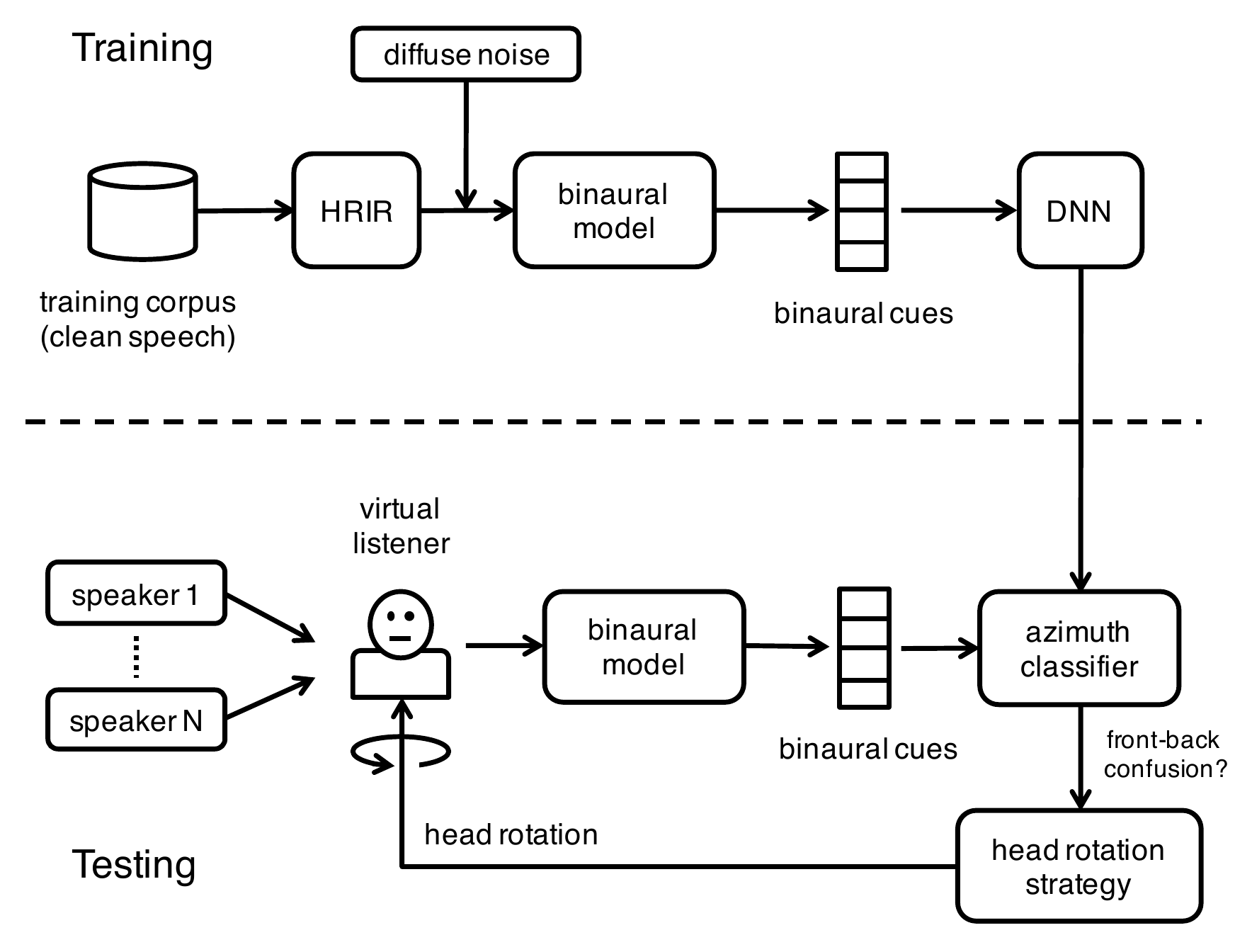}}
\caption{Schematic diagram of the proposed system, showing steps during training (top) and testing (bottom). During testing, sound mixtures consisting of several talkers are rendered in a virtual acoustic environment, in which a binaural receiver is moved in order to simulate the head rotation of a listener.}
\label{f:system}
\end{figure}

Figure~\ref{f:system} shows a schematic diagram of the proposed binaural sound localisation system in the full 360$\,\deg$ azimuth range. During training, clean speech signals were spatialised using \glspl{hrir}, and diffuse noise was added before being processed by a binaural model for feature extraction. The noisy binaural features were used to train \glspl{dnn} to learn the relationship between binaural cues and sound azimuths. During testing, sound mixtures consisting of several talkers are rendered in a virtual acoustic environment, in which a binaural receiver is moved in order to simulate the head rotation of a human listener. The output from the \gls{dnn} is combined with a head movement strategy to robustly localise multiple talkers in reverberant environments.

\subsection{Binaural feature extraction}
\label{ss:feature}

An auditory front-end was employed to analyse binaural ear signals with a bank of $32$ overlapping Gammatone filters, with centre frequencies uniformly spaced on the \gls{erb} scale between 80\,Hz and 8\,kHz~\cite{WangBrown2006}. Inner-hair-cell processing was approximated by half-wave rectification. No low-pass filtering was employed to simulate the loss of phase-locking at high frequencies as previous studies have shown that in general classifiers are able to exploit the high-frequency structure~\cite{MayvandeParKohlrausch11a}. Afterwards, the \gls{ccf} between the right and left ears was computed independently for each frequency band using overlapping frames of 20\,ms with a 10\,ms shift. The \gls{ccf} was further normalised by the auto-correlation value at lag zero~\cite{MayvandeParKohlrausch11a} and evaluated for time lags in the range of $\pm$1.1\,ms.


Two binaural features, \glspl{itd} and \glspl{ild}, are typically used in binaural localisation systems \cite{Blauert97}. The \gls{itd} is estimated as the lag corresponding to the maximum in the cross-correlation function. The \gls{ild} corresponds to the energy ratio between the left and right ears within the analysis window, expressed in dB. In this study, instead of estimating \gls{itd} the entire \gls{ccf} was used as localisation features. This approach was motivated by two observations. First, computation of \glspl{itd} involves a peak-picking operation which may not be robust in the presence of noise and reverberation.
Second, there are systematic changes in the \gls{ccf} with source azimuth (in particular, changes in the main peak with respect to its side peaks). Even in multi-source scenarios, these can be exploited by a suitable classifier. For signals sampled at 16\,kHz, the \gls{ccf} with a lag range of $\pm$1\,ms produced a 33-dimensional binaural feature space for each frequency band. This was supplemented by the \gls{ild}, forming a 34-dimensional (34D) feature vector.



%
\subsection{DNN localisation}
\label{ss:dnn}

\glspl{dnn} were used to map the 34D binaural feature set to corresponding azimuth angles. A separate DNN was trained for each of the 32 frequency bands. Employing frequency-dependent DNNs was found to be effective for localising simultaneous sound sources. Although simultaneous sources overlap in time, within a local time frame each frequency band is mostly dominated by a single source (Bregman's~\cite{Bre1990} notion of `exclusive allocation'). Hence, this allows training using single-source data and removes the need to include multi-source data for training.

The \gls{dnn} consists of an input layer, two hidden layers, and an output layer. The input layer contained 34 nodes and each node was assumed to be a Gaussian random variable with zero mean and unit variance. The 34D binaural feature inputs for each frequency band were Gaussian normalised, and white Gaussian noise (variance 0.4) was added to avoid overfitting, before being used as input to the \gls{dnn}. The hidden layers had sigmoid activation functions, and each layer contained 128 hidden nodes. The number of hidden nodes was heuristically selected -- more hidden nodes increased the computation time but did not improve localisation accuracy. The output layer contained 72 nodes corresponding to the 72 azimuth angles in the full 360$\deg$ azimuth range, with a 5$\deg$ step. A `softmax' activation function was applied at the output layer. The same \gls{dnn} architecture was used for all frequency bands and we did not optimise it for individual frequencies.

The neural network was initialised with a single hidden layer, and the number of hidden layers was gradually increased in later training phases. In each training phase, mini-batch gradient descent with a batch size of 128 was used, including a momentum term with the momentum rate set to 0.5. The initial learning rate was set to 1, which gradually decreased to 0.05 after 20 epochs. After the learning rate decreased to 0.05, it was held constant for a further 5 epochs. We also included a validation set and the training procedure was stopped earlier if no new lower error on the validation set could be achieved within the last 5 epochs. At the end of each training phase, an extra hidden layer was added between the last hidden layer and the output layer, and the training phase was repeated until the desired number of hidden layers was reached (two hidden layers in this study). 

Given the observed feature set $\vec{x}_{t,f}$ at time frame $t$ and frequency band $f$, the 72 `softmax' output values from the DNN for frequency band $f$ were considered as posterior probabilities $\mathcal{P}(k|\vec{x}_{t,f})$, where $k$ is the azimuth angle and $\sum_k \mathcal{P}(k|\vec{x}_{t,f}) = 1$. The posteriors were then integrated across frequency to yield the probability of azimuth $k$, given features of the entire frequency range at time $t$
\begin{equation}
\mathcal{P}(k|\vec{x}_t) =  \frac{P(k)\prod\nolimits_f  \mathcal{P}(k|\vec{x}_{t,f})}{\sum\nolimits_{k}P(k)\prod\nolimits_f  \mathcal{P}(k|\vec{x}_{t,f})}, 
\label{e:post_t}
\end{equation}
where $P(k)$ is the prior probability of each azimuth $k$. Assuming no prior knowledge of source positions and equal probabilities for all source directions, Eq.~\ref{e:post_t} becomes
\begin{equation}
\mathcal{P}(k|\vec{x}_t) =  \frac{\prod\nolimits_f  \mathcal{P}(k|\vec{x}_{t,f})}{\sum\nolimits_{k}\prod\nolimits_f  \mathcal{P}(k|\vec{x}_{t,f})}.
\end{equation}
Sound localisation was performed for a signal block consisting of $T$ time frames. Therefore the frame posteriors were further averaged across time to produce a posterior distribution $\mathcal{P}(k)$ of sound source activity
\begin{equation}
\mathcal{P}(k) = \frac{1}{T} \sum_t^{t+T-1} \mathcal{P}(k|\vec{x}_{t}).
\label{math:post}
\end{equation}
The target location was given by the azimuth $k$ that maximised $\mathcal{P}(k)$
\begin{equation}
\hat{k} = \argmax_k \mathcal{P}(k)
\end{equation}

\subsection{Localisation with head movements}
\label{ss:head}

\begin{figure}[!t]
%
%
%
\centerline{\includegraphics[width=\columnwidth]{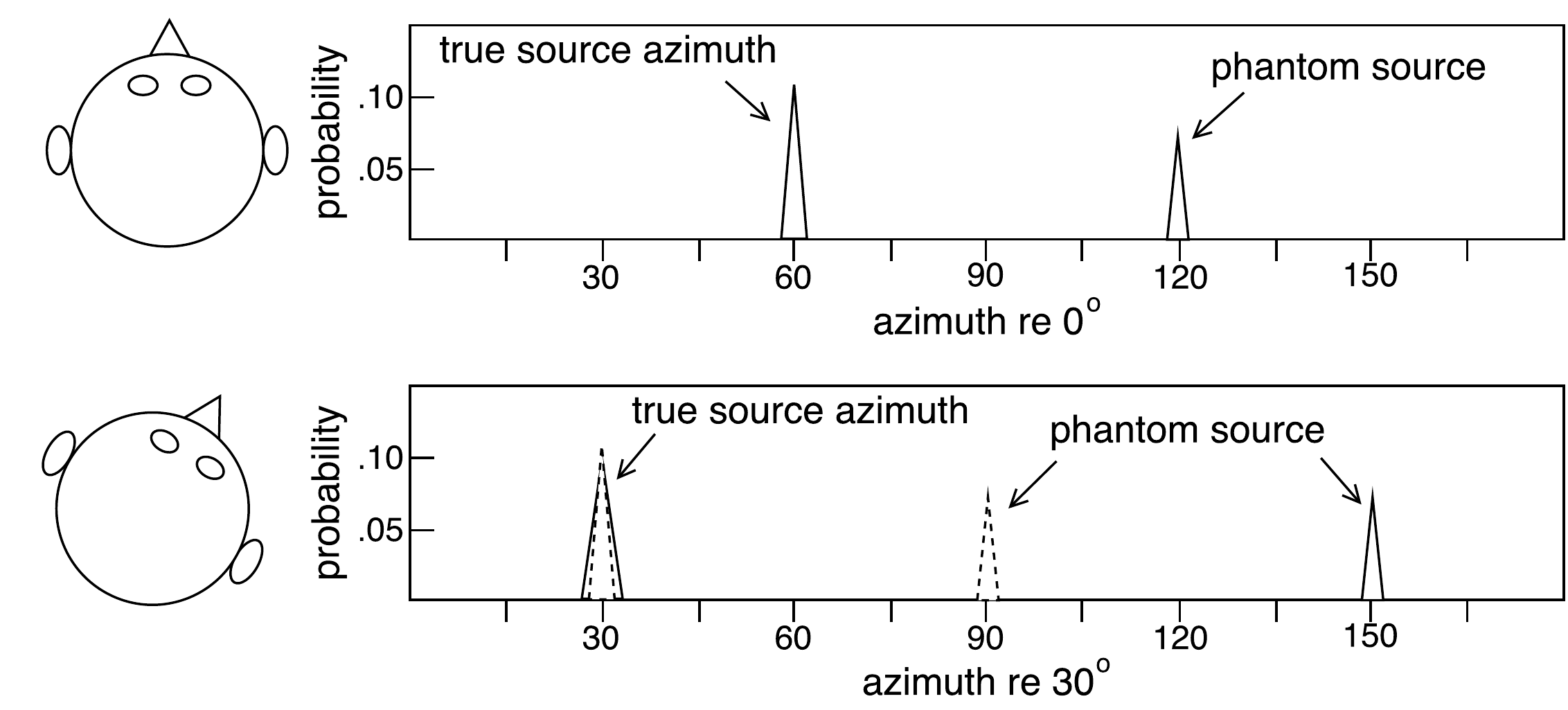}}
\caption{Illustration of the head movement strategy. Top: posterior probabilities where two candidate azimuths at 60$\deg$ and 120$\deg$ are identified. Bottom: after head rotation by 30$\deg$, only the azimuth candidate at 30$\deg$ agrees with the azimuth-shifted candidate from the first signal block (dotted line).}	
\label{f:HeadMovementStrategy}
\end{figure}

In order to reduce the number of front-back confusions, the proposed localisation model employs a hypothesis-driven feedback stage that triggers a head movement if the source location cannot be unambiguously estimated. A signal block is used to compute an initial posterior distribution of the source azimuth using the trained \glspl{dnn}. In an ideal situation, the local peaks in the posterior distribution correspond to the azimuths of true sources. However, due to the similarity of binaural features in the front and rear hemifields, \emph{phantom sources} may also become apparent as peaks in the azimuth posterior distribution. Such an ambiguous posterior distribution is shown in the top panel of Figure~\ref{f:HeadMovementStrategy}. In this case, a random head movement within the range of [$-$30$\deg$,\,30$\deg$] is triggered to solve the localisation confusion. Other possible strategies for head movement are discussed in \cite{MaEtAl2015}.

A second posterior distribution is computed for the signal block after the completion of the head movement. If a peak in the first posterior distribution corresponds to a true source position, then it will appear in the second posterior distribution and will be shifted by an amount corresponding to the angle of head rotation (assuming that sources are stationary before and after the head movement). On the other hand, if a peak is due to a phantom source, it will not occur in the second posterior distribution, as shown in the bottom panel of Figure~\ref{f:HeadMovementStrategy}. By exploiting this relationship, potential phantom source peaks are identified and eliminated from both posterior distributions. After the phantom sources have been removed, the two posterior distributions were averaged to further emphasise the local peaks corresponding to true sources. The most prominent peaks in the averaged posterior distribution were assumed to correspond to active source positions. Here the number of active sources was assumed to be known \emph{a priori}.


The proposed approach to exploiting head movements is based on late information fusion -- the information from the model predictions is integrated. This is in contrast to the approach in \cite{Braasch2013} which adopted early fusion at the feature level by averaging cross-correlation patterns across different head orientations. Late fusion is preferred here for a couple of reasons: i) the use of head rotation is not needed during model training and thus it is more straightforward to generate data for training robust localisation models (DNNs); ii) early feature fusion tends to lose information which can otherwise be exploited by the system. As a result, the proposed system is able to deal with overlapping sound sources in reverberant conditions, while the system reported in \cite{Braasch2013} was tested in anechoic conditions with a single source.


\section{Evaluation}
\label{s:eval}

\subsection{Binaural simulation}

Binaural audio signals were created by convolving monaural sounds with \glspl{hrir} or \glspl{brir}. For training, an anechoic \gls{hrir} catalog based on the \gls{kemar} head and torso simulator with pinnae~\cite{WierstorfGeierRaakeSpors11} was used for simulating the anechoic training signals. The \gls{hrir} catalog included impulse responses for the full 360$\,\deg$ azimuth range, allowing us to train localisation models for 72 azimuths between 0$\deg$ and 355$\deg$ with a 5$\deg$ step. The models were trained using only the anechoic \glspl{hrir} and were not retrained for any room conditions. See Section~\ref{ss:mct} for more details about training.

For evaluation, the Surrey \gls{brir} database~\cite{HummersoneMasonBrookes10} and a \gls{brir} set recorded at TU Berlin~\cite{MaEtAl2015} were used to reflect different reverberant room conditions. The Surrey database was recorded using a Cortex \gls{hats} and includes four room conditions with various amounts of reverberation. The loudspeakers were placed around the \gls{hats}  on an arc in the median plane, with a 1.5\,m radius between $\pm$90$\deg$ and measured at 5$\deg$ intervals. Table~\ref{t:rooms} lists the reverberation time (T$_{60}$) and the \gls{drr} of each room. The anechoic \glspl{hrir} used for training were also included to simulate an anechoic condition.

\begin{table}[thb]
\caption{Room characteristics of the Surrey \gls{brir} database~\cite{HummersoneMasonBrookes10}.}
\label{t:rooms}
\vspace{1mm}
\centering
\begin{tabular}{@{} l  c  c  c  c  @{}}
\hline\hline
 & Room A & Room B  & Room C & Room D \\
\hline
$\mathrm{T}_{60}$ (s) & 0.32 & 0.47 & 0.68 & 0.89  \\
$\mathrm{DRR}$ ($\deci\bel$) & 6.09 & 5.31 & 8.82 & 6.12 \\
\hline\hline
\end{tabular}
\end{table}

\begin{figure}[thb]
\centerline{\includegraphics[width=.7\columnwidth]{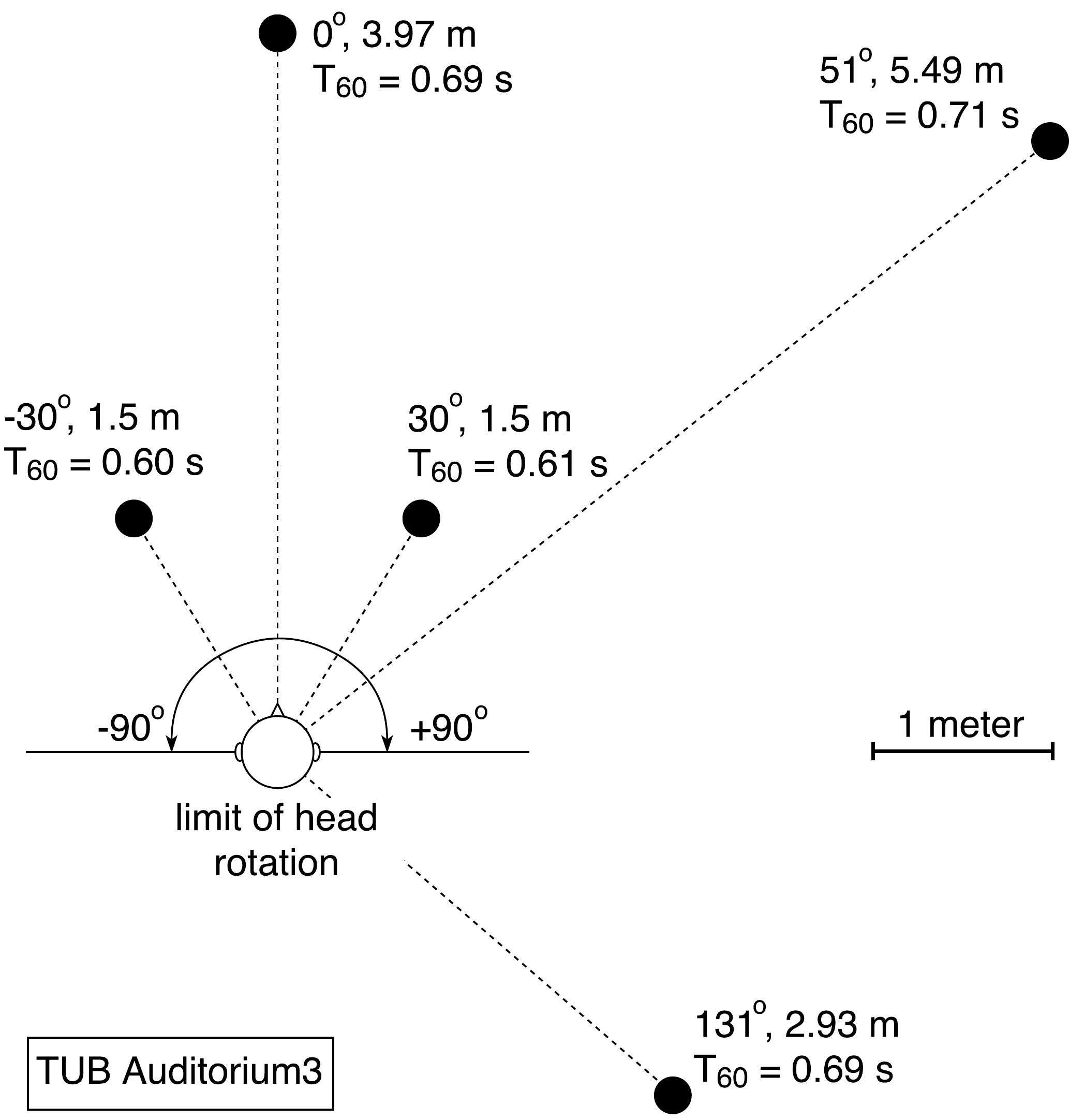}}
\caption{Schematic diagram of the TUB Auditorium3 configuration. The source distance, azimuth angle and respective T$_{60}$ time are shown for each source.}
\label{f:room3}
\end{figure}

A second set of \glspl{brir}, recorded in the ``Auditorium3'' room at TU Berlin\footnote{The BRIRs are freely available at \url{http://tinyurl.com/lt76yqs}}, was also included particularly for evaluating the benefit of head movements (Section~\ref{ss:expr_headmove}). The Auditorium3 room is a mid-size lecture room of dimensions 9.3\,m $\times$ 9\,m, with a trapezium shape and an estimated reverberation time T$_{60}$ of 0.7\,s. The \gls{brir} measurements were made for different head orientations ranging from $-$90$\deg$ to 90$\deg$ with an angular resolution of 1$\deg$. \glspl{brir} for six different source positions, including one in the rear hemifield, were recorded and five of them were selected for this study (two 0$\deg$ positions are available and the one at 1.5\,m away from the head was excluded for simplicity). The five selected source positions with respect to the dummy head are illustrated in Figure~\ref{f:room3}. 

Note that the anechoic \glspl{hrir} used for training and the Surrey \glspl{brir} were recorded using two different dummy heads (\gls{kemar} and Cortex \gls{hats}). We use data from two dummy heads because this study is concerned with sound localisation in the 360$\deg$ azimuth range; the Surrey \gls{hats} \glspl{hrir} catalog is only available for the frontal azimuth angles and therefore cannot be used to train the full 360$\deg$ localisation models. However, as the experiment results will show in Section IV, with \gls{mct} our proposed systems generalised well despite the \gls{hrir} mismatch between training and testing.

\begin{figure}[thb]
\centerline{\includegraphics[width=.7\columnwidth]{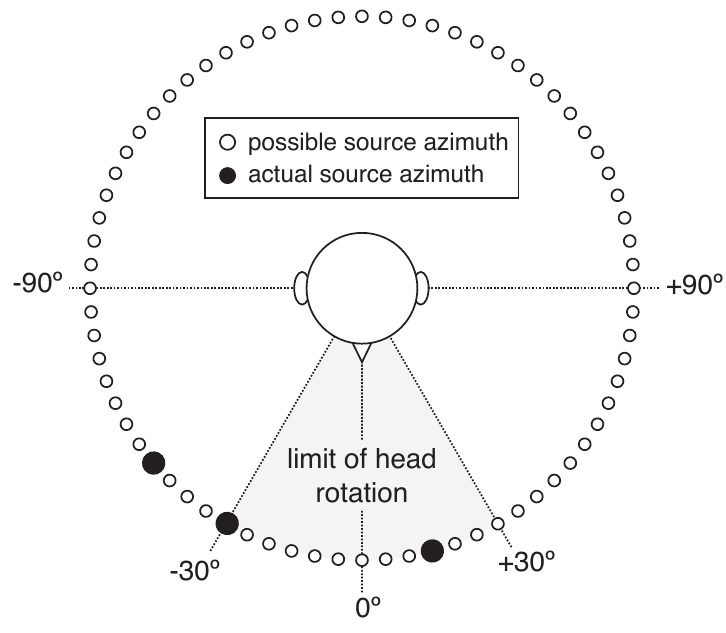}}
\caption{Schematic diagram of the Surrey BRIR room configuration. Actual source positions were always between $\pm$90$\deg$, but the system could report a source azimuth at any of 72 possible azimuths around the head (open circles). Black circles indicate actual source azimuths in a typical three-talker mixture (in this example, at $-$50$\deg$, $-$30$\deg$ and 15$\deg$). During testing, head movements were limited to the range $\left[-30\deg, 30\deg\right]$ as shown by the shaded area.}
\label{f:azimuth}
\end{figure}

Binaural mixtures of multiple competing sources were created by spatialising each source separately at the respective \gls{brir} sampling rate, before adding them together in each of the two binaural channels. In the Auditorium3 BRIRs there is varying distance between the listener position and different source positions. Furthermore there is a difference in impulse response amplitude level even for sources of the equal distance to the listener, likely due to the microphone response difference across recording sessions. To compensate the level difference a scaling factor was computed for each source position by averaging the maximum levels in the impulse responses between left and right ears. The scaling factors were used to adjust the level for each source before spatialisation. As a result the direct sound level of each source when mixed together was approximately the same. For the Surrey BRIR set the level difference did not exist and thus this preprocessing was not applied. The spatialised signals were finally resampled to 16\,kHz for training and testing.

\subsection{Head movement simulation}

For the Surrey \glspl{brir}, head movements were simulated by computing source azimuths relative to the head orientation, and loading corresponding \glspl{brir} for the relative source azimuths. Such simulation is only approximate for the reverberant room conditions because the Surrey \gls{brir} database was measured by moving loudspeakers around a fixed dummy head. With the Auditorium3 \glspl{brir}, more realistic head movements were simulated by loading the corresponding BRIR for a desired head orientation. For all experiments, head movements were limited to the range of $\pm30\,\deg$.

\subsection{Multi-conditional training}
\label{ss:mct}

The proposed systems assumed no prior knowledge of room conditions. The localisation models were trained using only anechoic \glspl{hrir} with added diffuse noise, and no reverberant \glspl{brir} were used during training.

Previous studies~\cite{MayvandeParKohlrausch11a,WoodruffWang12,MayvandeParKohlrausch13,MayMaBrown2015} have shown that \gls{mct} features can increase the robustness of localisation systems in reverberant multi-source conditions. 
Binaural \gls{mct} features were created by mixing a target signal at a specified azimuth with diffuse noise at various \glspl{snr}. The diffuse noise is the sum of 72 uncorrelated, white Gaussian noise sources, each of which was spatialised across the full 360$\,\deg$ azimuth range in steps of 5$\,\deg$. Both the directional target signals and the diffuse noise were created using the same anechoic \gls{hrir} recorded using a \gls{kemar} dummy head~\cite{WierstorfGeierRaakeSpors11}. This approach was used in preference to adding reverberation during training, since previous studies (e.g., \cite{WoodruffWang12}) suggested that it was more likely to generalise well across a wide range of reverberant test conditions.

The training material consisted of speech sentences from the {TIMIT} database~\cite{Gar1993}. A set of 30 sentences was randomly selected for each of the 72 azimuth locations. For each spatialised training sentence, the anechoic signal was corrupted with diffuse noise at three \glspl{snr} (20, 10 and 0\,dB \gls{snr}). The corresponding binaural features (\glspl{itd}, \gls{ccf}, and \glspl{ild}) were then extracted. Only those features for which the \textit{a priori} \gls{snr} between the target and the diffuse noise exceeded $-$5\,dB were used for training. This negative \gls{snr} criterion ensured that the multi-modal clusters in the binaural feature space at higher frequencies, which are caused by periodic ambiguities in the cross-correlation analysis, were properly captured.

\subsection{Experimental setup}


The {GRID} corpus~\cite{GridCorpus} was used to create three evaluation sets of 50 acoustic mixtures which consisted of one, two or three simultaneous talkers, respectively. Each GRID sentence is approximately 1.5\,s long and was spoken by one of 34 native British-English talkers. The sentences were normalised to the same \gls{rms} value prior to spatialisation. For the two-talker and three-talker mixtures, the additional azimuth directions were randomly selected from the same azimuth range while ensuring an angular distance of at least 10$\deg$ between all sources. Each evaluation set included 50 acoustic mixtures which were kept the same for all the evaluated azimuths and room conditions in order to ensure any performance difference was due to test conditions rather than signal variation. Since the duration of each GRID sentence was different, and there was silence of various lengths at the beginning of each sentence, the central 1\,s segment of each sentence was selected for evaluation.

Note that although the models were trained and evaluated using speech signals, our systems are not intended to localise only speech sources. Therefore a frequency range from 80\,Hz to 8\,kHz was selected for the signals sampled at 16\,kHz. Our previous studies~\cite{MayMaBrown2015, MaEtAl2015dnn} also show that 32 Gammatone filters (see Section~\ref{ss:feature}) provide a good tradeoff between frequency resolutions and computational cost. As the evaluation included localisation of up to three overlapping talkers, using too few filters would result in insufficient frequency resolution to reliably localise multiple talkers.

The baseline system was a state-of-the-art localisation system~\cite{MayMaBrown2015} that modelled both \glspl{itd} and \glspl{ild} features within a \gls{gmm} framework. As in \cite{MayMaBrown2015}, the \gls{gmm} modelled the binaural features using 16 Gaussian components and diagonal covariance matrices for each azimuth and each frequency band. The \gls{gmm} parameters were initialised by 15 iterations of the $k$-means clustering algorithm and further refined using $5$ iterations of the \gls{em} algorithm. The second localisation model was the proposed DNN system using the \gls{ccf} and ILD features. Each DNN employed four layers including two hidden layers each consisting of 128 hidden nodes (see Section~\ref{ss:dnn}).

Both localisation systems were evaluated using different training strategies (clean training and MCT), various localisation feature sets (ITD, ILD and \gls{ccf}), and with or without head movements. When no head movement was employed, the source azimuths were estimated using the entire 1\,s segment from each acoustic mixture. If head movement was used, the 1\,s segment was divided into two 0.5\,s long blocks and the second block was provided to the system after completion of a head movement. Therefore in both conditions the same signal duration was used for localisation. 

The gross localisation accuracy was measured by comparing true source azimuths with estimated azimuths. The number of active sources $N$ was assumed to be known \textit{a priori} and the $N$ azimuths for which the posterior probabilities were the largest were selected as the estimated azimuths. Localisation of a source was considered accurate if the estimated azimuth was less than or equal to 5$\deg$ away from the true source azimuth:
\begin{equation}
\text{LocAcc} = \frac{N_{\text{dist}(\phi, \hat{\phi}) \le \theta}}{N}
\end{equation}
where $\text{dist}(.)$ is the angular distance between two azimuths, $\phi$ is the true source azimuth, $\hat{\phi}$ is the estimated azimuth, and $\theta$ is the threshold in degrees (5$\deg$ in this study). This metric is preferred to RMS errors because our study is concerned with full 360$\deg$ localisation, and localisation errors in degrees are often large due to front-back confusions.


\section{Results and Discussion}
\label{s:expr}



\begin{table*}[thb]
\caption{Gross localisation accuracy in $\%$ for various sets of \glspl{brir} when localising one, two and three competing talkers in the frontal hemifield only and in the full 360$\deg$ range.} 
\label{t:loc_acc_mct}
\centering
\begin{tabular}{@{} l  c  c  c c c  c c c   c c c   c c c   c c c   c @{}}
\hline\hline
Hemi- & & & \multicolumn{3}{c}{Anechoic}  & \multicolumn{3}{c}{Room A} & \multicolumn{3}{c}{Room B} & \multicolumn{3}{c}{Room C} & \multicolumn{3}{c}{Room D} & \\
\cline{4-18}
filed & \multirow{-2}{*}{Model}	&\multirow{-2}{*}{MCT} &1 &2 &3  &1 &2 &3  &1 &2 &3  &1 &2 &3  &1 &2 &3  &\multirow{-2}{*}{Avg.} \\
\hline
& 					& no		&100   &99.0  &90.5   &84.0  &63.1  &52.8   &81.5  &59.8  &51.8   &100   &82.5  &65.5   &88.2  &61.2  &53.5   &75.6 \\
&\multirow{-2}{*}{GMM}	& yes	&100   &99.9  &98.7   &99.2  &97.1  &90.7   &100   &97.7  &91.6   &100   &99.3  &96.5   &100   &98.4  &91.5   &97.4 \\
\cline{2-19}
&					&no		&100   &100   &99.6   &100   &99.2  &92.2   &100   &99.0  &90.4   &100   &99.9  &96.7   &99.9  &98.7  &91.1   &97.8 \\
\multirow{-4}{*}{Frontal} &\multirow{-2}{*}{DNN}	&yes		&100   &100   &99.7   &100   &99.5  &96.3   &100   &99.7  &96.2   &100   &99.9  &98.2   &100   &99.6  &95.3   &99.0 \\
\hline
& 					& no		&100   &97.1  &82.6   &82.6  &48.9  &30.7   &65.6  &38.3  &25.3   &98.4  &70.3  &50.2   &77.2  &46.3  &30.0   &62.9 \\
&\multirow{-2}{*}{GMM}	& yes	&100   &100   &97.8   &99.0  &94.2  &80.7   &97.0  &89.0  &77.6   &100   &97.6  &88.7   &97.3  &90.6  &79.0   &92.6 \\
\cline{2-19}
&					&no		&100   &100   &97.4   &100   &87.0  &68.4   &94.5  &79.0  &63.9   &97.7  &92.5  &78.9   &94.4  &83.4  &67.9   &87.0 \\
\multirow{-4}{*}{360$\deg$} &\multirow{-2}{*}{DNN}	&yes		&100   &100   &98.6   &99.7  &97.3  &87.9   &97.2  &93.7  &86.7   &100   &97.3  &90.2   &97.3  &94.0  &85.0   &95.0 \\
\hline\hline
\multicolumn{19}{@{}l}{The models were trained using either clean training or the MCT method.}
\end{tabular}
\end{table*}  

\subsection{Influence of \gls{mct}}
\label{ss:expr_mct}

The first experiment investigated the impact of \gls{mct} on the localisation accuracy of the proposed systems. Two scenarios were considered: i) sound localisation was restricted to the frontal hemifield so that the systems estimated source azimuths within the range [$-$90$\deg$,\,90$\deg$]; ii) the systems were not informed that the sources lay only in the frontal hemifield and were free to report the azimuth in the full 360$\deg$ azimuth range. In the second scenario front-back confusions could occur.

Table~\ref{t:loc_acc_mct} lists gross localisation accuracies of all the systems evaluated using various \gls{brir} sets from the Surrey database. First consider the scenario of localisation in the frontal hemifield. For the \gls{gmm} baseline system, the \gls{mct} approach substantially improved the robustness across all conditions, with an average localisation accuracy of 97.4\% compared to only 75.6\% using clean training. The improvement with \gls{mct} was particularly large in multi-talker scenarios and in the presence of room reverberation. For the \gls{dnn} system, the improvement with MCT over clean training was not as large as that for the GMM system and is only observed in the multi-talker scenarios. The limited improvement is partly because with clean training the performance of the DNN system is already very robust in most conditions, with an average accuracy of 97.8\%, which is already better than the GMM system with MCT. This suggests that when localisation was restricted to the frontal hemifield, the DNN can effectively extract cues from the clean CCF-ILD features that are robust in the presence of reverberation.

Consider the case of full 360$\deg$ localisation, the scenario is more challenging and front-back errors could occur. The GMM system with clean training failed to localise the talkers accurately, with error rates greater than 50\% when localising multiple simultaneous talkers. The DNN system with clean training was substantially more robust than the GMM system, but the performance also decreased significantly when multiple talkers were present. The benefit of the MCT method became more apparent for both systems in this scenario -- the average localisation accuracy was increased from 62.9\% to 92.6\% for the GMM system and from 87\% to 95\% for the DNN system. Across all the room conditions the largest benefits were observed in room B where the direct-to-reverberant ratio was the lowest, and in room D where the reverberation time T$_{60}$ was the longest.

\begin{figure*}[thb]
\centerline{\includegraphics[trim=5cm 0cm 4.5cm 0cm,clip=true,width=.9\textwidth]{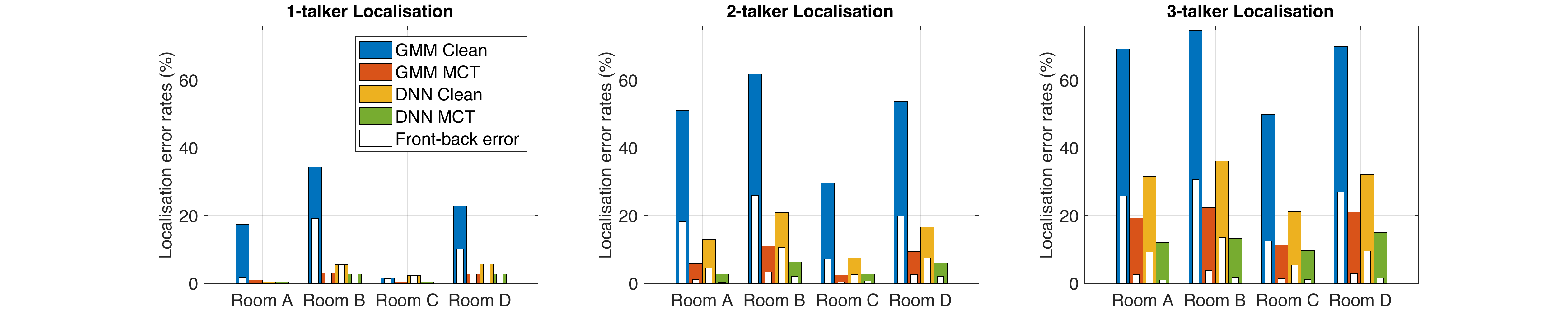}}
\caption{Localisation error rates produced by various systems using either clean training or MCT. Localisation was performed in the full 360$\deg$ range, so that front-back errors could occur, as shown by the white bars for each system. No head movement strategy was employed.}
\label{f:loc_errors_mct}
\end{figure*}

Errors made in 360$\deg$ localisation could be due to front-back confusion as well as interference caused by reverberation and overlapping talkers. Figure~\ref{f:loc_errors_mct} shows errors made by both the GMM and the DNN systems using either clean training or MCT in different room conditions. The errors due to front-back confusions were indicated by white bars for each system. Here a localisation error is considered to be a front-back confusion when the estimated azimuth is within $\pm20$ degrees of the azimuth that would produce the same ITDs in the rear hemifield. It is clear that front-back confusions contributed a large portion of localisation errors for both systems, in particular when clean training was used. When the MCT method was used, not only the errors due to interference of reverberation and overlapping talkers (non-white bar portion in Figure~\ref{f:loc_errors_mct}) were greatly reduced, but also the systems produced substantially fewer front-back errors (white bars in Figure~\ref{f:loc_errors_mct}). As will be discussed in the next section, without head movements the main cues distinguishing between front-back azimuth pairs lie in the combination of inteaural level and time differences (or ITD-related features such as the cross-correlation function). MCT provides the training stage with better regularisation of the features, which is able to improve the generalisation of the learned models and better discriminate the front-back confusing azimuths.

It is also worth noting that the training and testing stages used HRTFs collected with different dummy heads (the KEMAR was used for training and the HATS was used for testing). However, with MCT the localisation accuracy in the anechoic condition for localising one or two sources was 100\%, which suggests that MCT also reduced the sensitivity to mismatches of the receiver.

\subsection{Contribution of the ILD cue}
\label{ss:expr_ild}

\begin{table*}[thb]
\caption{Gross localisation accuracy in $\%$ using various feature sets for localising one, two and three competing talkers in the full 360$\deg$ range.} 
\label{t:loc_acc_ild}
\centering
\begin{tabular}{@{} c  c  c c c  c c c   c c c  c c c  c c c  c@{}}
\hline\hline
&  & \multicolumn{3}{c}{Anechoic}  & \multicolumn{3}{c}{Room A} & \multicolumn{3}{c}{Room B} & \multicolumn{3}{c}{Room C} & \multicolumn{3}{c}{Room D} & \\
\cline{3-17}
\multirow{-2}{*}{Model} & \multirow{-2}{*}{Feature}	 &1 &2 &3  &1 &2 &3  &1 &2 &3  &1 &2 &3  &1 &2 &3  &\multirow{-2}{*}{Avg.} \\
\hline
	& ITD				&100   &99.8  &96.2   &99.2  &81.6  &67.7   &91.4  &76.6  &64.9   &97.2  &89.4  &76.6   &89.1  &76.6  &65.8   &84.8 \\
	& \textbf{ITD-ILD}		&100   &100   &97.8   &99.0  &94.2  &80.7   &97.0  &89.0  &77.6   &100   &97.6  &88.7   &97.3  &90.6  &79.0   &\textbf{92.6} \\
\multirow{-3}{*}{GMM} &CCF-ILD	&100   &100   &98.4   &100   &87.2  &73.9   &92.1  &81.7  &71.5   &99.9  &93.8  &81.6   &92.6  &83.2  &72.3   &88.5 \\
\hline
	& CCF				&100   &100   &99.0   &99.8  &95.8  &86.7   &91.8  &89.5  &83.7   &98.3  &95.8  &89.0   &91.6  &87.8  &80.8   &92.7 \\
\multirow{-2}{*}{DNN} & \textbf{CCF-ILD}		&100   &100   &98.6   &99.7  &97.3  &87.9   &97.2  &93.7  &86.7  &100   &97.3  &90.2   &97.3  &94.0  &85.0   &\textbf{95.0} \\
\hline\hline
\multicolumn{17}{@{}l}{The models were trained using the MCT method. The best feature set for each system is marked in bold font.}
\end{tabular}
\end{table*}

The second experiment investigated the influence of different localisation features, in particular the contribution of the ILD cue. Table~\ref{t:loc_acc_ild} lists the gross localisation accuracies using various feature sets. Here all models were trained using the MCT method and the active head movement strategy was not applied. When ILDs were not used, the GMM performance using just ITDs suffered greatly in reverberant rooms and when localising overlapping talkers; the average localisation accuracy decreased from 92.6\% to 84.8\%. The performance drop was particularly pronounced in rooms B and D, where the reverberation was strong. For the DNN system, excluding the ILDs also decreased the localisation performance but the performance drop was more moderate, with the average accuracy reduced from 95\% to 92.7\%. The DNN system using the \gls{ccf} feature exhibited more robustness in the reverberant multi-talker conditions than the GMM system using the ITD feature. As previously discussed, computation of the ITD involved a peak-picking operation that could be less reliable in challenging conditions, and the systematic changes in the \gls{ccf} with the source azimuth provided richer information that could be exploited by the DNN.

\begin{figure*}[thb]
\centerline{\includegraphics[trim=5cm 0cm 4.5cm 0cm,clip=true,width=.9\textwidth]{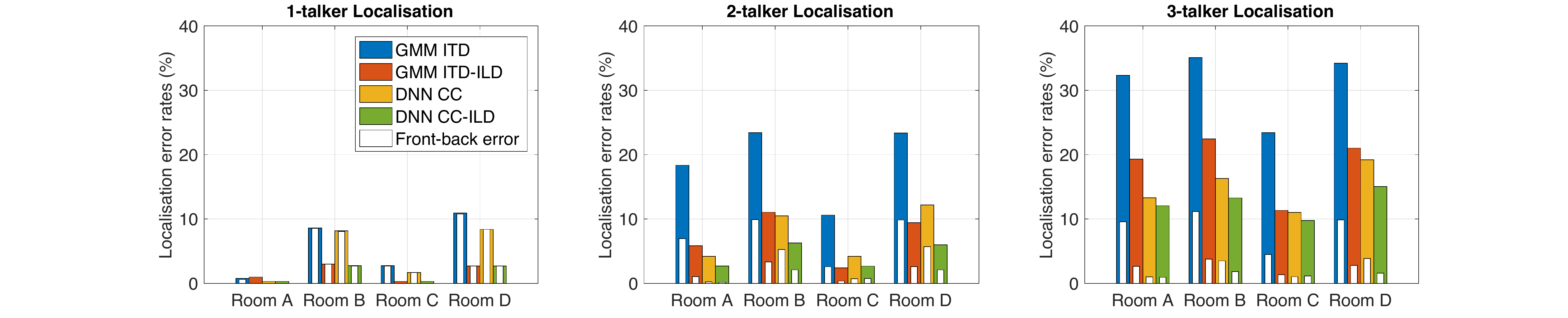}}
\caption{Comparison of localisation error rates produced by various systems using different spatial features. Localisation was not restricted in the frontal hemifield so that front-back errors can occur, as indicated by the white bars for each system. No head movement strategy was employed.}
\label{f:loc_errors_ild}
\end{figure*}

When ILDs were not used, the localisation errors were largely due to an increased number of front-back errors as suggested by Figure~\ref{f:loc_errors_ild}. For single-talker localisation in rooms B and D, without using ILDs almost all the errors made by the systems were front-back errors. When ILDs were used, the number of front-back errors were greatly reduced in all conditions. This suggests that the ILD cue plays a major role in solving the front-back confusions. ITDs or ILDs alone may appear more symmetric between the front and back hemifields, but together with ILDs they create the necessary asymmetries (due to the KEMAR head with pinnae) for the models to learn the differences between front and back azimuths.


Table~\ref{t:loc_acc_ild} also lists localisation results of the GMM system when using the same CCF-ILD feature set as used by the DNN system. The GMM failed to extract the systematic structure in the \gls{ccf} spanning multiple feature dimensions, most likely due to its inferior ability to model correlated features. The average localisation accuracy is only 88.5\% compared to 95\% for the DNN system, and again it suffered the most in more reverberant conditions such as rooms B and D.

\subsection{Benefit of the head movement strategy}
\label{ss:expr_headmove}

Table~\ref{t:loc_acc_head} lists the gross localisation accuracies with or without head movement. All systems were trained using the MCT method and employed the respective best performing features (\textit{GMM ITD-ILD} and \textit{DNN CCF-ILD}).

\begin{table*}[thb]
\caption{Gross localisation accuracies in $\%$ with or without the head movement when localising one, two and three competing talkers in the full 360$\deg$ azimuth range.} 
\label{t:loc_acc_head}
\centering
\begin{tabular}{@{}c  c   c c c  c c c   c c c   c c c   c c c   c@{}}
\hline\hline
				   & Head & \multicolumn{3}{c}{Anechoic}  & \multicolumn{3}{c}{Room A} & \multicolumn{3}{c}{Room B} & \multicolumn{3}{c}{Room C} & \multicolumn{3}{c}{Room D} & \\
\cline{3-17}
\multirow{-2}{*}{Model} & move &1 &2 &3  &1 &2 &3  &1 &2 &3  &1 &2 &3  &1 &2 &3  &\multirow{-2}{*}{Avg.} \\
\hline
					& no		&100   &100   &97.8  &99.0  &94.2  &80.7  &97.0  &89.0  &77.6  &100   &97.6  &88.7  &97.3  &90.6  &79.0  &92.6 \\
\multirow{-2}{*}{GMM}	& yes	&100   &100   &97.5  &100   &97.3  &83.4  &99.8  &93.1  &79.9  &99.9  &99.3  &90.8  &99.9  &93.0  &79.5  &94.2 \\
\hline
					&no		&100   &100   &98.6  &99.7  &97.3  &87.9  &97.2  &93.7  &86.7  &100   &97.3  &90.2  &97.3  &94.0  &85.0  &95.0 \\
\multirow{-2}{*}{DNN} 	&yes		&100   &100   &98.4  &100   &99.2  &90.0  &99.8  &96.1  &86.9  &100   &99.0  &91.6  &99.5  &94.7  &84.7  &96.0 \\
\hline\hline
\multicolumn{17}{@{}l}{All systems were trained using the MCT method.}
\end{tabular}
\end{table*}

\begin{figure*}[thb]
\centerline{\includegraphics[trim=4.2cm 0cm 4.2cm 0cm,clip=true,width=.9\textwidth]{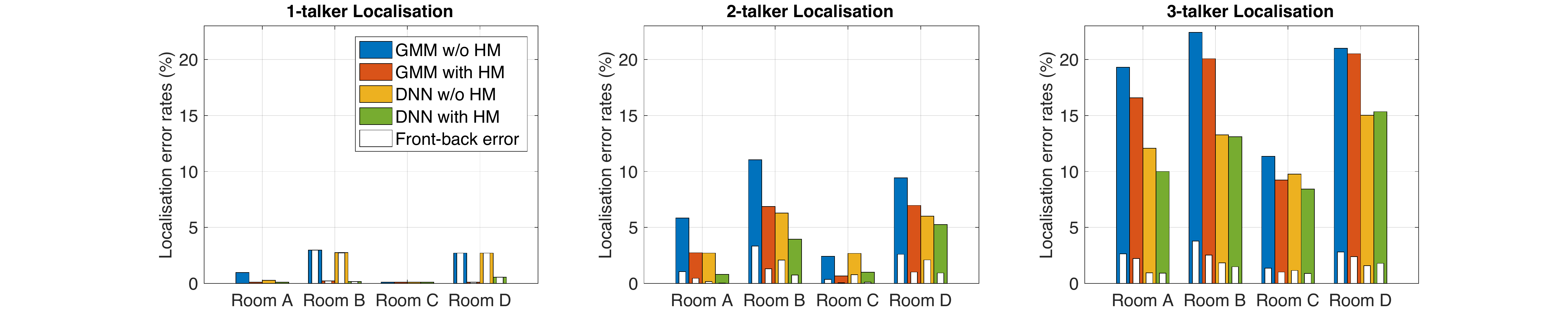}}
\caption{Localisation error rates produced by various systems with or without head movement when localising one, two or three overlapping talkers. Localisation was performed in the 360$\deg$ azimuth range so that front-back errors can occur, as indicated by the white bars for each system.}
\label{f:loc_errors_head}
\end{figure*}


Both the GMM and DNN systems benefitted from the use of head movements. It is clear from Figure~\ref{f:loc_errors_head} that the localisation errors were almost entirely due to front-back confusions in one-talker localisation. By exploiting the head movement, the systems managed to reduce most of the front-back errors and achieved near 100\% localisation accuracies. In two- or three-talker localisation, the number of front-back errors was also reduced with the use of head movements. When overlapping talkers were present, the systems produced many localisation errors other than front-back errors, due to the partial evidence available to localise each talker. By removing most front-back errors, the systems were able to further improve the accuracy of localising overlapping sound sources.

\begin{figure*}[thb]
\centerline{\includegraphics[trim=4cm 0.5cm 4cm 1cm,clip=true,width=.8\textwidth]{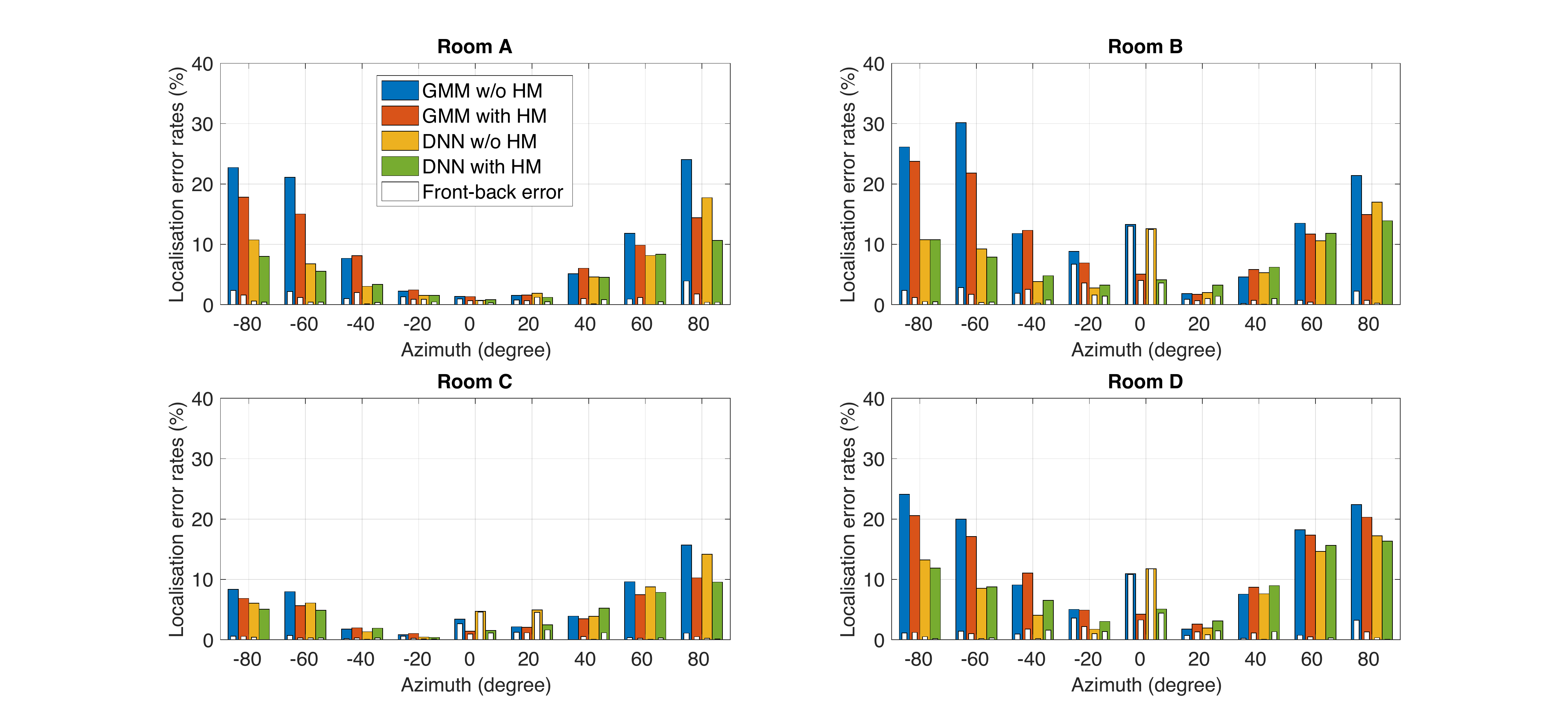}}
\caption{Localisation error rates produced by various systems with or without head movement, as a function of the azimuth. The histogram bin width is 20$\deg$. Here the error rates were averaged across the 1-, 2- and 3-talker localisation tasks. Localisation was performed in the full 360$\deg$ azimuth range so that front-back errors can occur, as indicated by the white bars for each system.}
\label{f:loc_errors_head_az}
\end{figure*}

Figure~\ref{f:loc_errors_head_az} shows the localisation error rates as a function of the azimuth. The error rates here were averaged across the 1-, 2- and 3-talker localisation tasks. Across most room conditions, sound localisation was generally more reliable at more central locations than at lateral source locations. This is particularly the case for the GMM system, as shown in Figure~\ref{f:loc_errors_head_az}, where the localisation error rates for sources at the sides were above 20\% even in the least reverberant Room A. It is also clear from Figure~\ref{f:loc_errors_head_az} (white bars) that localisation errors were mostly not due to front-back confusions at lateral azimuths, and in this case the proposed DNN system outperformed the GMM system significantly.

At the central azimuths, on the other hand, almost all the localisation errors were due to front-back confusions. It is noticeable that in more reverberant conditions (such as Rooms B and D), the error rates at the central azimuths \mbox{[$-$10$\deg$,\,10$\deg$]} were particularly high due to front-back errors for both the GMM and the DNN systems when head movement was not used. The front-back errors were concentrated at central azimuths, probably because binaural features (interaural time and level differences) were less discriminative between 0$\deg$ and 180$\deg$ than between the more lateral azimuth pairs.


\begin{figure*}[thb]
\centerline{\includegraphics[trim=4.2cm 0cm 4.2cm 0cm,clip=true,width=.9\textwidth]{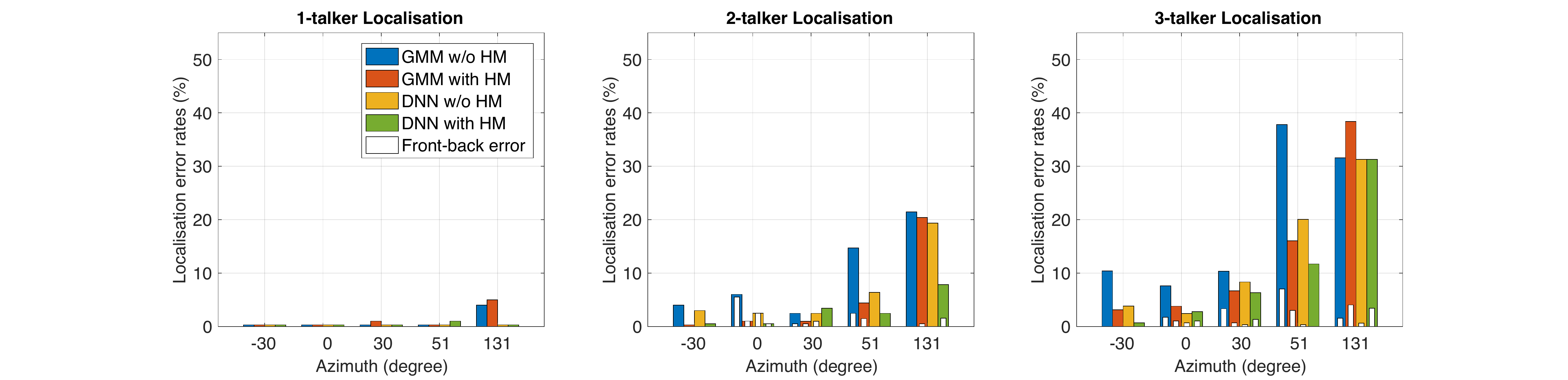}}
\caption{Localisation error rates produced by various systems as a function of the azimuth for the Auditorium3 task. Localisation was performed in the full 360$\deg$ azimuth range so that front-back errors can occur, as indicated by the white bars for each system.}
\label{f:loc_errors_head_TUB}
\end{figure*}

Finally, Figure~\ref{f:loc_errors_head_TUB} shows the localisation error rates using the Auditorium3 BRIRs in which head movements were more accurately simulated by loading the corresponding BRIR for a given head orientation. Overall the DNN systems significantly outperformed the GMM systems. For single-source localisation the DNN system achieved near 100\% localisation accuracy for all source locations including the one at 131$\deg$ in the rear hemifield. The GMM system produced about 5\% error rate for rear source but performed well for the other locations. For two- and three-source localisation, both GMM and DNN systems benefitted from head movements across most azimuth locations. For the GMM system the benefit is particularly pronounced for the source at 51$\deg$, with localisation reduced from 14\% to 4\% in two-source localisation and from 36\% to 14\% in two-source localisation. The rear source at 131$\deg$ appeared to be difficult to localise for the GMM system even with head movement, with 20\% error rate in two-source localisation. The DNN system with head movements was able to reduce the error rate for the rear source at 131$\deg$ to 8\%.

In general the performance of the models for the 51$\deg$ and 131$\deg$ locations is worse than the other source locations when there are multiple sources present at the same time. This is more likely due to the nature of the room acoustics at these locations, e.g. they are further away from the listener and closer to walls. When the sources are overlapping with each other, there are less glimpses left for localisation of each source and with stronger reverberation the sources at 51$\deg$ and 131$\deg$ became more difficult to localise.


\section{Conclusion}
\label{s:conc}


This paper presented a machine-hearing framework that combines \glspl{dnn} and head movements for robust localisation of multiple sources in reverberant conditions. Since simultaneous talkers were located in a full 360$\deg$ azimuth range, front-back confusions occurred. Compared to a GMM-based system, the proposed \gls{dnn} system was able to exploit the rich information provided by the entire \gls{ccf}, and thus substantially reduced localisation errors. The \gls{mct} method was effective in combatting reverberation, and allowed anechoic signals to be used for training a robust localisation model that generalised well to unseen reverberant conditions and to mismatched artificial heads used in training and testing conditions. It was also found that the inclusion of \glspl{ild} was necessary for reducing front-back confusions in reverberant rooms. The use of head rotation further increased the robustness of the proposed system, with an average localisation accuracy of 96\% under acoustic scenarios where up to three competing talkers and room reverberation were present.


In the current study, the use of DNNs allowed higher-dimensional feature vectors to be exploited for localisation, in comparison with previous studies~\cite{MayvandeParKohlrausch11a,WoodruffWang12,MayMaBrown2015}. This could be carried further, by exploiting additional context within the DNN either in the time or the frequency dimension. Moreover, it is possible to complement the features used here with other binaural features, e.g. a measure of interaural coherence~\cite{Fal2004}, as well as monaural localisation cues, which are known to be important for judgment of elevation angles~\cite{AsanoEtAl1990,ZakarauskasEtAl1993}. Visual features might also be combined with acoustic features in order to achieve audio-visual source localisation. 

The proposed system has been realised in a real world human-robot interaction scenario. The azimuth posterior distributions from the DNN for each processing block were temporally smoothed using a leaky integrator and head rotation was triggered if a front-back confusion was detected in the integrated posterior distribution. Audio signals acquired during head rotation were not processed. Such a scheme can be more practical for a robotic platform as head rotation often produces self-noise which makes the audio unusable.

One limitation of the current systems is that the number of active sources is assumed to be known \textit{a priori}. This can be improved by including a source number estimator that is either learned from the azimuth posterior distribution output by the DNN, or provided directly as an output node in the DNN. The current study only deals with the situation where sound sources are static. Future studies will relax this constraint and address the localisation and tracking of moving sound sources within the \gls{dnn} framework.

\section*{Acknowledgements}

This work was supported by the European Union FP7 project {\sc TWO!EARS} (http://www.twoears.eu) under grant agreement No.~618075.

\ifCLASSOPTIONcaptionsoff
  \newpage
\fi



%

\bibliographystyle{IEEEtran}
\bibliography{journal_abrv_short,twoears}

%

\begin{IEEEbiography}[{\includegraphics[width=1in,height=1.25in,clip,keepaspectratio]{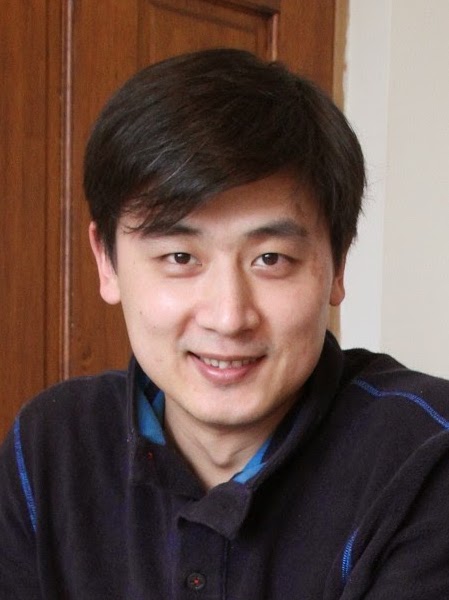}}]{Ning Ma}
obtained a MSc (distinction) in Advanced Computer Science in 2003 and a PhD in hearing-inspired approaches to automatic speech recognition in 2008, both from the University of Sheffield. He has been a visiting research scientist at the University of Washington, Seattle, and a Research Fellow at the MRC Institute of Hearing Research, working on auditory scene analysis with cochlear implants. Since 2015 he is a Research Fellow at the University of Sheffield, working on computational hearing. His research interests include robust automatic speech recognition, computational auditory scene analysis, and hearing impairment. He has authored or coauthored over 40 papers in these areas.
\end{IEEEbiography}


\begin{IEEEbiography}[{\includegraphics[width=1in,height=1.25in,clip,keepaspectratio]{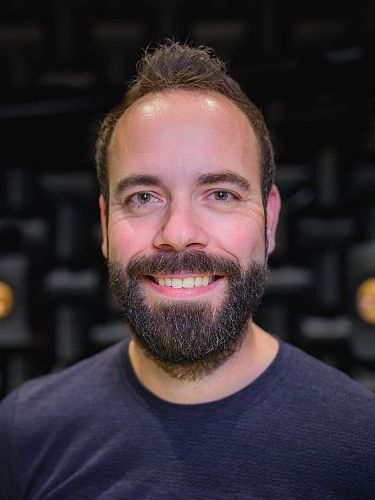}}]{Tobias May}
studied hearing technology and audiology and received his M.Sc. degree from the University of Oldenburg, Germany in 2007. In 2012 he received his binational Ph.D. degree from the University of Oldenburg, Germany in collaboration with the Eindhoven University of Technology, The Netherlands. Since 2013, he has been employed at the Department of Electrical Engineering at the Technical University of Denmark, first as a Postdoctoral Researcher (2013-2017), and now as an Assistant Professor (since 2017). His research interests include computational auditory scene analysis, binaural signal processing, noise-robust speaker identification and hearing aid processing.
\end{IEEEbiography}

\begin{IEEEbiography}[{\includegraphics[width=1in,height=1.25in,clip,keepaspectratio]{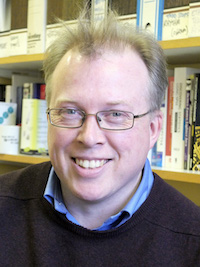}}]{Guy J. Brown}
obtained a BSc (Hons) Applied Science from Sheffield City Polytechnic in 1984 and a PhD in Computer Science from the University of Sheffield in 1992. He was appointed to a Chair in the Department of Computer Science, University of Sheffield, in 2013. He has held visiting appointments at LIMSI-CNRS (France), Ohio State University (USA), Helsinki University of Technology (Finland) and ATR (Japan). His research interests include computational auditory scene analysis, speech perception, hearing impairment and acoustic monitoring for medical applications. He has authored more than 100 papers and is the co-editor (with Prof. DeLiang Wang) of the IEEE book ``Computational auditory scene analysis: principles, algorithms and applications''.
\end{IEEEbiography}




\end{document}